\documentstyle [12pt] {article}

\def\be{\begin{equation}}
\def\ee{\end{equation}}

\begin{document}
\title {\large \bf  Bicovariant Differential Geometry of the Quantum Group
$SL_h(2)$ }
\author {V. Karimipour}
\maketitle
\begin {center}
{\it Institute for studies in Theoretical Physics and Mathematics ({\bf IPM })
\\ P.O.Box 19395-5746 Tehran, Iran \\
and \\
 Department of Physics , Sharif University of Technology\\
P.O.Box 11365-9161 Tehran, Iran\\}.

\end{center}
\vspace {10 mm}
\begin {abstract}

There are only two quantum group structures on the space of two by two
unimodular
matrices, these are the $SL_q(2)$ and the $SL_h(2)$ [9-13] quantum groups.
One can not construct a differential geometry on $ SL_q(2) $ , which at the
same time is bicovariant , has three generators , and satisfies the
Liebnitz rule. We show that such a differential geometry exists for the
quantum group $ SL_h(2) $ and derive all of it's properties.

\end{abstract}

\pagebreak
\section {Introduction}

One can look at a quantum group in two different ways:
As the quantization of the algebra of functions on a poisson-Lie group [1,2]
, or as a quantum automorphism group acting on a non-commuting space [3].

The former point of view is best suited for revealing the mathematical unity of
integrable models [1,4].
If one demands that, the process of quantization, preserves some natural
properties of the poisson Lie groups , then it can be shown that the
resulting quantum group will be unique. This is the quantization in the sence
of Drinfeld [1].

The second point of view is however best suited for the application  of quantum
group in formulating physical theories like quantum mechanics and
field theory ,
on non-commuting space time [5,7]. This second approach is however not as
rigid as
the first one, and one can imagine quantum spaces with a variety of defining
relations,which will lead to different quantum groups.
It's therefore natural to seek for criteria that restricts the set of quantum
spaces and hence
quantum groups. One such criteria is the demand of the existence of a central
determinant
in the quantum matrix group. \\ Fortunately, in the space of two by two
matrices this
demand, restricts the quantum groups to only two classes [8,9]. These are the
$SL_q (2)$ and the $SL_h(2)$ quantum groups.

The latter was first introduced in [10] for the case of $h=1$, where it was
called
the Jordanian deformation of $sl(2)$. The continuous  parameter $h$ was then
introduced
in [11]. Later the quantized universal enveloping algebra $U_h (sl(2)$ was
found
[12] , and finally, the quantum planes [13,14] and the structure of the quantum
De Rham
Complexes on the plane $R_h (2)$ and the group $SL_h(2)$ were constructed.
[13].

In the light of the non-comutative Geometry [15] and it's possible relevance to
physics [16], the explicit construction of all the objects characterizing the
differential geometry of a quantum group is important. This is true especially
for the simplest quantum group $SL_q(2)$ and $SL_h(2)$. Following the
poineering work of Woronowics [17],
Bicovariant differential Geometries have been constructed
for many compact and non compact q-groups [18,19]. In this paper we construct
the Bi-covariant differential geometry of $SL_h (2)$. \\  The structure of this
paper is as follows:
In section 2, we describe the $SL_h(2)$ quantum group and the quantum De Rham
Complex on it . In section 3, we describe our basic strategy for completing
the structure of the bicovariant differential geometry of $SL_h(2)$.
Section 4,gives the explicit results and formulas for various differential
geometric objects pertaining to $SL_h(2)$. Finally, in section 5, we use the
alge
bra of vector fields
on $SL_h(2)$ we derive a new deformations of the $sl(2)$ algebra, admitting a
truely quadratic casimiar.

\section{The Quantum Group $SL_h(2)$ and it's quantum De Rham Complex}

The quantum group $SL_h (2)$ is defined as a unital Hopf algebra generated
by the elements of a quantum matrix

$$ T= \left(  \begin{array}{ll}
a&b\\c&d
\end{array}
\right) $$
subject to the following relations.

$$[c,a]=h c^2   \hskip 3cm [b,a]=h(1- a^2)$$
\be[c,d]=h c^2    \hskip 3cm [b,d]=h(1- d^2)\ee
$$[a,d]=h (a-d)c  \hskip 3cm [c,b]=h (ac+cd)$$
with the h-determinant $ D_h(T) = ad - bc - h ac = 1 $.
These relations can also be obtained from the $RTT=TTR$ relations  for the
following $R$-matrix.

$$ R= \left( \begin{array}{llll}
1&h&-h&h^2\\ &1&0&h\\ & &1&-h\\
 & & &1 \end{array} \right) $$

Note that the h-determinant  $D_h (T)$ is  grouplike
,i.e: ( for $[t_{ij}$,$t'_{kl}]=0$) : $D_h (TT') = D_h (T) D_h (T')$.
The Hopf structure of $SL_h(2)$ is defined as usual:
$$ \Delta T = T\otimes T$$
$$\varepsilon (T)=1.$$
$$ S(T)= \left( \begin{array}{ll}
d-hc&-b+ha-hd+h^2c\\
c& \ \ \ \  a+hc   \end{array}
\right) $$
\vskip 2cm
One can also check that: if, $T \epsilon SL_h(2)$ and $T' \epsilon SL_{h'}
(2)$,
then $TT' \epsilon SL_{h+h'}(2)$ and $S(T) \epsilon SL_{-h} (2)$. This algebra
has also a two parametric extension $SL_ {h,h'} (2)$ , which has been
studied in ref. [20].
For the dual Algebra, that's $U_h (sl(2))$ see [12]. It's also interesting to
note that this quantum
group can be understood as a "Quantum automorphism group" [13-14] of a pair of
noncommutative linear spaces defined by the relations:
$$xy - yx = -h y^2$$
$$ \xi^2 = - h \xi \eta \hskip 1cm \eta^2 = 0 \hskip 1cm \xi \eta + \eta \xi =
0 $$ where $\xi = \hat{d} x$ and $\eta= \hat{d} y$
In Ref. [13], the complete structure of the differential graded algebra of
these quantum planes together with the structure of the quantum de Rham Complex
of the group itself has been studied. Here we give a brief account of this part
of [13].
One starts from the ansatz
\be R_{12} \hat{d} T_1 T_2 = T_2 \hat{d} T_1 R_{12}\ee
the consistency of which with the defining relations (1), is due to a nice
property of the $R$--matrix, namely
\be R_{12} R_{21} = R_{21} R_{12} = 1\ee
 where $ R_{21} = P R_{12} P $  and $ P $ is the permutation matrix .
 Using this property one finds from  (2) that:
\be R_{12} ^{-1}\hat{d} T_2 T_1 = T_1 \hat{d} T_2 R_{12} ^{-1}\ee
 combination of (2) and (4) leads to
\be \hat{d} (R_{12} T_1 T_2 - T_2 T_1 R_{12})=0\ee
\ \  which proves the consistency of the ansatz (2).
If one then defines the left invarant one-forms as:
 \be \Omega =S(T)\hat{d} T\ee
 \ \ one will obtain:
\be \Omega _1 T_2 R_{21} = T_2 R_{21} \Omega _1\ee
\  \be \Omega _1 R_{12} \Omega_2 R_{21} + R_{21} \Omega _2 R_{21} \Omega _1 =
0\ee  \be \hat{d} \Omega + \Omega \wedge \Omega =0 \ee
In [13], the explicit relations are derived for the matrix $\Omega$ in the
form
\be \Omega= \left( \begin{array}{ll} \bf
{{w_+ + w_-} \over 2}& \ \ \bf u\\
\ \ \bf v&\bf {{w_+ - w_-} \over 2}
\end{array}
\right) \ee

However the relations of [13] are true for $GL_h (2)$. To obtain the
corresponding relations for $SL_h (2)$ one demands that
$$\hat{d} (1)=\hat{d} (ad-bc-hac)=0$$

For the $ SL-q(2) $ quantum group one can not impose such a relation , since
the determinant is not central in the differential algebra. It is only central
in the algebra itself. However  one can easily verify that in the $ SL_h(2)
$ quantum group the determinant is central in the  whole differential algebra.
A simple calculation  shows that this puts the following
restriction on the left invariant one forms:
$${\bf w_+} = 2h {\bf v}$$
which reduces the independent number of one-forms to three as expected.
In terms of the independ forms ${\bf w_- ,u}$ and ${\bf v}$, the results of
solution of
(7) and (8) are as follow: From (7):
\be{\bf w_-} a = a {\bf w_-} +2ha{\bf v} \hskip 2.5cm {\bf v}a=a{\bf v}\ee
\be\ \ \ \ {\bf w_-} b =b{\bf w_-}  -2hb {\bf v} \hskip 2.5cm {\bf v}b=b{\bf
v}+2ha{\bf v}\ee
\be{\bf w_-} c =c{\bf w_-}  +2hc{\bf v} \hskip 2.5cm  {\bf v}c=c{\bf v}\ee
\be\ \ \ \ {\bf w_-} d =d{\bf w_-}  -2hd{\bf v} \hskip 2.5cm  {\bf v}d=d{\bf
v}+2hc{\bf v}\ee
\be {\bf u}a=a {\bf u} -ha ({\bf w_-} +h{\bf v})\ee
\be {\bf u}b=b{\bf u} -2ha({\bf u}-h^2 {\bf v}) +hb({\bf w_-} -h{\bf v})\ee
\be {\bf u}c=c{\bf u} -hc({\bf w_-} +h{\bf v})\ee
\be {\bf u}d=d{\bf u}-2hc({\bf u}-h^2 {\bf v})+hd({\bf w_-} -h{\bf v})\ee

{}From (8):

$${\bf v} \wedge {\bf v } = {\bf w_-} \wedge {\bf w_-} = [{ \bf w_- ,v }]_+
= 0 $$ \be  [\ {\bf u , v} ]_+ =  h  [ {\bf v , w _- }]_- \hskip 2cm  [\ \bf {u
,w_-}]_+  = 2h {\bf
[ u, v]_-} \ee  $${\bf u} \wedge{\bf u } =  h  [  {\bf w_-} , {\bf u} - 2 h^2
{\bf v } ]_-    $$
where $ {\bf   [  \alpha  , \beta  ]_{\pm} } \equiv {\bf \alpha}  \wedge {\bf
\beta } \pm {\bf \beta} \wedge {\bf \alpha }$  .
Note that the 2-forms ${\bf u}\wedge {\bf v} , {\bf w_- } \wedge {\bf v}$ and
${\bf
w_-}\wedge {\bf  u}$ span the space of 2-forms. Taking this into account eq.(9
), leads to: $$ d{\bf w_- +2u}\wedge {\bf v} + 2 h { \bf w_-} \wedge {\bf
v} = 0 $$
\be d {\bf u +w_-} \wedge {\bf u} - 4h^2 {\bf w_-} \wedge {\bf v} -2h {\bf u}
\wedge {\bf v}=0 \ee $$d{\bf v - w_-} \wedge {\bf v}=0 $$
{}From equation (6) one obtains $  {\hat d} T = T \Omega $ , inserting from
equation ( 10 ) the form of $ \Omega$ and taking into account the constraint
$
{\bf w_+} =2h{\bf v}$
gives the
definition of the differentials $\hat{d} a,....
\hat{d} d$ in terms of the left invaraiant  one forms :
$$\hskip 4cm \hat{d} a=(b+ha){\bf v} +{a\over 2}{\bf w_-} \hskip 4cm (20-a) $$
$$\hskip 4cm  \hat{d} b=a{\bf u}+hb{\bf v}-{b\over 2}{\bf w_-}\hskip 4cm
(20-b)$$
$$\hskip 4cm \hat{d} c=(hc+d){\bf v}+{c\over 2}{\bf w_-}\hskip 4cm (20-c)$$
$$\hskip 4cm \hat{d} d=c{\bf u}+hd{\bf v}-{d\over 2}{\bf w_-}\hskip 4cm
(20-d)$$
In the next section we combine the above results and the general construction
of [17] to construct a bicovariant differential geometry on $SL_h (2)$.

\section {Bicovariant Differential Geometry of $SL_h (2)$}

Let A be a Hopf algebra ( here $Fun_h(SL(2))$, briefly $SL_h(2)$ ). $\Gamma$ an
A-bimodule, i.e: the space of quantum one forms (here span $\{\hat{d}
a,...\hat{d} d\}$ ) , \ \
$_{inv} \Gamma$ the space of left invariant 1-forms (here span $\{ w_i = {\bf
w_1, u,v } \}$ ),and finally
$\Gamma_{inv}$, the space of right invariant 1-forms. \\  Let $\Delta
_L:\Gamma \rightarrow
A \otimes \Gamma$ and $\Delta _R: \Gamma \rightarrow \Gamma \otimes A$ be the
quantum analogue of the pullback of one forms under left and right
multiplication
 of the group:
\be\Delta _L (adb)=\Delta (a) (id\otimes d)\Delta b\ee
\be \Delta _R (adb)=\Delta (a) (d\otimes id)\Delta b\ee
Then the following relations hold [17].
\be\Delta _L {\bf w_i}=1 \otimes {\bf w_i}\ee
\be\Delta _R {\bf w_i=w_j} \otimes M_i ^j \ee

The elements $M_i ^j$ then define the adjoint representation of the quantum
group.\\ As shown by Woronowicz [17] there exists functionals $f^i _j$ and
$\chi _i$ such that:
\be {\bf w^i} a= (id\otimes f^i _j) \Delta (a) {\bf w^j}\ee
\be d a= (id\otimes \chi _i) \Delta (a) {\bf w^i}\ee
 Here $f^i _j $ and $ \chi _i$ are linear maps from A to the field over which A
is defined.
The functionals $f^i _j$'s characterize the noncomutativity of the algebra, and
$\chi_i $'s
are the quantum analog of left-invariant vector fields.One can also prove that:
\be d{\bf w^i} +C^i_{jk} {\bf w^j} \otimes {\bf w^k} = 0 \ee
where $C^k _{ij} = \chi _j (M_i^k)$. The wedge product is constructed using the
quantum analog of the flip automorphism: $\Lambda ({\bf w^i} \otimes {\bf
w^j}) = \Lambda ^{ij}_{kl} {\bf w^k} \otimes {\bf w^l}$:
\be {\bf w^i} \wedge {\bf w^j} ={\bf w^i} \otimes {\bf w^j} -\Lambda ^{ij}
_{kl} {\bf w^k} \otimes {\bf w^l}\ee Where
\be \Lambda ^{ij} _{kl}= f^i _l (M_k ^j) \ee

For brevity we do not give the explicit form of the elements $ M_k^i $.They are
need only in the clculation of the coefficients $ \Lambda^{ij}_{kl} $. The
result of this calculation will be given later.

The Lie algebra of vector fields is defined by the following relations:
\be \chi _i \chi _j -\Lambda ^{kl} _{ij} \chi _k \chi _l=C_{ij} ^k \chi _k
\ee  for the Hopf structure of $\chi _i, f^j _i$ and $M_i ^j$ see [17,19].
Having at our disposal the structure of the quantum De Rham Complex of $SL_h
(2)$
(Section (2) )  we next apply the above formalism to it, and obtain, all the
other objects characterizing the geometry of $SL_h (2)$.

\subsection { Vector fields and the functionals $f^i _j$}

{}From eqs- ((20-b)-(20-e)) and (26) we obtain the following  evaluations, for
$\chi _{\bf u},\chi _{\bf v}$ and $\chi_-$.
$$ < \chi _{\bf u} ,T> =\left( \begin{array}{ll}
0&1\\0&0 \end{array} \right) \hskip 1cm <\chi _{\bf v},T>= \left(
\begin{array}{ll}
h&0\\1&h \end{array} \right) \hskip 1cm <\chi_- ,T>={1\over 2} \left(
\begin{array}{ll} 1&0\\0&-1 \end{array} \right)$$

Similarly from eqs. (18) and (25) we obtain:
$$<f_- ^- ,T>= \left( \begin{array}{ll} 1&0\\0&1  \end{array} \right)
\hskip 1cm <f_{\bf v} ^- ,T>=<-2f_- ^{\bf u} , T>= \left( \begin{array}{ll}
2h&0\\  0&-2h  \end{array} \right) $$
$$ <f_{\bf v} ^{\bf v} ,T>=\left( \begin{array}{ll} 1&2h\\ 0&1 \end{array}
\right)
\hskip 1cm <f_{\bf u} ^{\bf u} ,T>= \left( \begin{array}{ll} 1&-2h\\0&1
\end{array} \right)$$
$$ <f_{\bf v} ^{\bf u} ,T>= h^2 \left( \begin{array}{ll} -1&2h\\0&-1
\end{array} \right) $$ and:
$<f_j  ^i ,T>=0 \ \ \ $ for$\ \ \ \    f_j ^i =f_{\bf u} ^- , f_- ^{\bf v}
,f_{\bf u} ^{\bf v}$

\subsection{Tensor product realization of the wedge}

A straightforward but lenghty calculation (using eq.(29)) gives the
coefficents $\Lambda ^{ij} _{kl}$, as follow:

$$\Lambda ^{\alpha \beta} _{\beta \alpha}=1 \hskip 2cm \forall \alpha , \beta$$
$$\Lambda ^{\bf uu} _{\bf -u}=-\Lambda ^{\bf uu} _{\bf u-}=\Lambda ^{\bf uv}
_{\bf v-}=-\Lambda^{\bf vu}_{\bf -v}=-2h$$
\be\Lambda ^{\bf v-}_{\bf vv}=\Lambda ^{\bf -u}_{\bf uv}=-\Lambda ^{\bf
-v}_{\bf vv}=-\Lambda ^{\bf u-} _{\bf vu}=-4h\ee
$$\Lambda ^{\bf uu}_{\bf uv}=\Lambda ^{\bf uu}_{\bf vu}=\Lambda ^{\bf uv}_{\bf
vv}=\Lambda ^{\bf u-}_{\bf v-}
=\Lambda ^{\bf vu}_{\bf vv}=\Lambda ^{\bf -u}_{\bf -v}= -4h^2$$
$$\Lambda ^{\bf u-}_{\bf vv}= -\Lambda^{\bf -u}_{\bf vv}= -8h^3$$
and all others equal to zero.

This then gives via eq.(28) the following tensor product realization for the
wedge product.

$${\bf u}\wedge {\bf u}=2h  \left( {\bf w_-} \otimes {\bf u-u}\otimes {\bf w_-
}+2h({\bf u}\otimes {\bf v+v }\otimes {\bf u})  \right)$$
$${\bf u}\wedge {\bf v=u}\otimes {\bf v-v}\otimes {\bf u}+2h {\bf v}\otimes
{\bf w_-} +4h^2 {\bf v}\otimes {\bf v}$$
$${\bf u}\wedge {\bf w_-} ={\bf u}\otimes {\bf w_- -w_-} \otimes {\bf u} -4h
{\bf v}\otimes {\bf u} +4h^2({\bf v}\otimes {\bf w_- }+2h {\bf v}\otimes {\bf
v})$$
$${\bf v}\wedge {\bf u=v}\otimes {\bf u-u}\otimes {\bf v}-2h({\bf w_- }\otimes
{\bf v}-2h{\bf v}\otimes {\bf v})$$
$${\bf v}\wedge {\bf v}=0$$
$${\bf v}\wedge {\bf w_-}={\bf v}\otimes {\bf w_- -w_- }\otimes {\bf v} +4h
{\bf v}\otimes {\bf v}$$
$${\bf w_-} \wedge {\bf u=w_-} \otimes {\bf u-u}\otimes {\bf w_-} +4h{\bf
u}\otimes {\bf v} +4h^2 ({\bf w_- }\otimes {\bf v}-2h{\bf v}\otimes {\bf v})$$
$${\bf w_- }\wedge {\bf v=w_- }\otimes {\bf v-v}\otimes {\bf w_-} -4h{\bf
v}\otimes {\bf v}$$
$${\bf w_-} \wedge {\bf w_-} = 0$$
One can now check the associativity of the wedge product, the commutation
relations (19) and the
validity of Liebnitz rule: i.e : $d({\bf \alpha}\wedge {\bf \beta} )=(d
{\bf \alpha}
)\wedge {\bf \beta} + (-1)^ {deg \alpha } {\bf \alpha}\wedge  d {\bf \beta}
,\forall {\bf \alpha}  ,{\bf \beta} \in A\oplus \Gamma$
\subsection {The h-Lie algebra of the vector Fields}.

To obtain the structure constants $C^k _{ij}$ we insert the above
realization of the wedge product in equation ( 20 ) and then use (27)
.The result is:

$$ C^{\bf u}_{- {\bf u}}= -C^{\bf u} _{{\bf u} -}=C^{\bf v }_{{\bf v}
-}=-C^{\bf v} _{-{\bf v}} = {1\over 2}C^- _{\bf {uv}}={-1\over 2}C^- _{\bf
{vu}} =1$$.
$$C^{\bf u} _{{\bf u v }}=C^{\bf u} _{{\bf v u}}=C^- _{\bf {v-}}=C^- _{\bf
{-v}}={1\over 2}C^{\bf v} _{{\bf v v}} = 2h$$

and all others equal to zero.

Eq.(30) will then give the following h-Lie algebra for the vector fields:

$$[\chi _-,\chi _{\bf u}]= (1-2h\chi _{\bf u})\chi _{\bf u}$$
\be[\chi _-,\chi _{\bf v}]= -(1-2h\chi _{\bf u})(\chi _{\bf v} +2h\chi _-)\ee
$$[\chi _{\bf u} ,\chi _{\bf v}]=-2(1-2h\chi _{\bf u})(h\chi _{\bf u} -\chi
_-)$$

One can check the Jacobi identity for this deformed Lie algebra .
 Obviously
this algebra goes to the Lie algebra of $sl(2)$ in the limit $h\rightarrow 0$.

\subsection{Lie derivative of Left-invariant forms}

The Lie derivative of one forms along left invariant vector fields is defined
as $$L_{\chi _{i}}{\bf w_j} = (id \otimes \chi _i)\Delta _R {\bf w_j}$$
from which we obtain:

$$L_{\chi _{\bf u}} \left( \begin{array}{l} \bf
u\\ \bf v\\ \bf w_- \end{array} \right)=\left( \begin{array}{l} {\bf
w_-} +2h{\bf v}\\ \ \ \  0\\ \ \ \ -2 {\bf v} \end{array} \right) \hskip 2cm
L_{\chi_{\bf v}} \left( \begin{array}{l}
 {\bf u}\\ \bf v\\ \bf w_- \end{array} \right)=\left( \begin{array}{l}
\ \  2h{\bf u}\\ {\bf -w_-} + 4 h{\bf v}\\  2{\bf u}+2h{\bf w_-} \end{array}
\right)$$

$$L_{\chi_{-}} \left( \begin{array}{l} \bf
u\\ \bf v\\ \bf w_- \end{array} \right)= \left( \begin{array}{l} {\bf
-u}\\ {\bf v}\\2h{\bf v} \end{array} \right) $$

It's easy to see that these Lie derivatives, actually represent the h-Lie
algebra (32).i.e:

$$L_{\chi _{i}} L_{\chi _{j}}- \Lambda ^{kl}_{ij} L_{\chi _{k}} L_{\chi _{l}}=
C_{ij} ^k L_{\chi _{k}}$$

\section{ A new deformation of $sl_2$ Lie algebra}

As it stands eqs. (32) define a deformation of the Lie algebra of $sl(2)$.
However the form of these equatios can be put into a more symmetrical form by
the following redefinitions. Set:

$$J_0 \equiv \chi _- \hskip 2.5cm J_+ \equiv \chi _{\bf u} \hskip 2.5cm J_-
\equiv \chi _{\bf v} +h^2\chi _{\bf u}$$.

Then after some symmetrizations one gets:

$$[J_0 ,J_+]={1\over 2}(KJ_+ +J_+K)$$
\be[J_0 ,J_-]={-1\over 2} (KJ_- +J_- K)\ee
$$[J_+ ,J_-]= KJ_0 +J_0K$$
where $K=1-2h J_+$. This algebra is then a new deformation of the $sl(2)$ Lie
algebra.

The matrices
$$ J_0= \left( \begin{array}{ll}
{1\over 2}& 0\\0& {-1\over 2}
\end{array}
\right) \hskip 1cm
 J_+ = \left( \begin{array}{ll}
0&1\\0&0
\end{array}
\right)   \hskip 1cm
 J_- = \left( \begin{array}{ll}
h&h^2\\1&h
\end{array}
\right) $$
form a two-dimensional representation of this algebra.

It 's interesting to note that this algebra has indeed a Casimiar operator
given by:

\be C=J_+ J_- +J_- J_+ +2J_0 ^2 +{1\over 2} K^2\ee
 \ Note that all the deformation is effected by the element $ K $ which is
liniear in the generators.
One can also derive the commutation relations of $K$ with the other
elements:
$$ [ J_+ , K ] = 0 $$
\be [ J_0 , K ] = -h ( K J_+ + J_+ K ) \ee
$$ [ J_- , K ] = 2h ( K J_0 + J_0 K ) $$

If one then treats $K$ as an independent generator, the relations (33) and (35
) define a deformation of the Lie algebra of $gl(2)$.

It may be interesting to study the representation of the algebra (33) and to
see
if as in the $sl_q (2)$ case, it parallels those of the classical Lie algebra
$sl(2)$ for generic values of $h$ and if it shows pecularities for some
specific values of the deformation parameter.

\vskip 2cm
{\large {\bf Acknowledgements}} I would like to thank all my coleagues in the
physics department of {\bf IPM}, for interesting discussions. I also express my
sincere thanks to Miss. S. Daver Panah for typing the manuscript. \newpage
{\Large {\bf References}}
\begin{enumerate}
\item{1} V.G. Drinfeld, Quantum Groups, proc. Inter. Congr. Math. (Berkeley)
Vol 1, Academic Press, New York, 1986, pp.798-820
\item{2} L.A. Takhtajan in M.L.Ge and B.H.Zhao (eds), Introduction to Quantum
Groups
and Integrable Massive Models of Quantum. Field Theory,World Scientific,
Singapore, 1991.
\item{3} Yu. I. Manin, Quantum Groups and non-commutative Geometry, Preprint
CRM-1561, Montreal 1988
\item{4} L.D. Faddeev, INtegrable Models in 1+1 dimensional quantum Fields
theory
(Les Houches Lectures 1982), Elsevier, Amsterdam.
\item{5} I. Ya Aref'eva, I.V. Volovich, Mod. Phys. Lett. {\bf A6} (1991) 893;
phys. Lett {\bf B 264} (1991) 62; I.Ya. Aref'eva, G.E. Arutyunov, J. Geometry
and physics. {\bf 11} (1993)
\item{6} M. Ubriaco, Phys. Lett. {\bf A 163} (1992) 1
\item{7} J. Schwenk, J. Wess, Phys. Lett. {\bf B 291} (1992) 273
\item{8} P. Schupp, P. Watts, B. Zumino, Lett. Math. Phys. {\bf 25}, 139-147
(1992)
\item{9} M. Dubois-Viollette, G. Launer, Phys. Lett. {\bf B 245}, 175 (1990)
\item{10} E.E. Demidov etall. prog. theor. Phys. Supp {\bf 102}, 203-218 (1990)
\item{11} S. Zakrzewski; Lett. Math. Phys. {\bf 22}, 287-289 (1991)
\item{12} C. H. Ohn; Lett. Math. Phys. {\bf 25}, 85-88 (1992)
\item{13} V. Karimipour; "Quantum De Rham Complexes associated with $SL_h(2)$"
Lett. Math. Phys. in press
\item{14} B.A. Kupershmidt  J.Phys.A Math.Gen. 25, L1239 (1992)
\item{15} A. Connes "Essays on Physics and Non-Commutative Geometry"
The Interface of Mathematics and Particle Physics, Clarendon Press Oxford
(1990)
\item{16} A. Connes and J. Lott; Nucl. Phys. {\bf B 18} (proc. Suppl.) (1990).
\item{17} S.L. Woronowics; Commun. Math. Phys. {\bf 122}, 125 (1989)
\item{18} V. Carrow-Watumura, et all. Commun. Math. Phys. {bf 142}, 605 (1991)
\item{19} P.Aschieri and L. Castellani; Int. Jour. Mod. Phys. {\bf A 8} 10
(1993);
L. Castellani; Phys. Lett. {\bf B 292}; 93 (1992); Phys. Lett. {\bf B 279}, 291
(1992)
\item{20} A. Aghamohammadi; Mod. Phys. Lett. {\bf A8}, 2607 (1993)

\newpage
\vfil\break
 \end{enumerate}
\end{document}